# Covalent Functionalization and Passivation of Exfoliated Black Phosphorus via Aryl Diazonium Chemistry

Christopher R. Ryder[1], Joshua D. Wood[1], Spencer A. Wells[1], Yang Yang[2], Deep Jariwala[1], Tobin J. Marks[1,2], George C. Schatz[2], and Mark C. Hersam[1,2*]

**Abstract**

Functionalization of atomically thin nanomaterials enables the tailoring of their chemical, optical, and electronic properties. Exfoliated black phosphorus – a layered two-dimensional semiconductor exhibiting favorable charge carrier mobility, tunable bandgap, and highly anisotropic properties – is chemically reactive and degrades rapidly in ambient conditions. In contrast, here we show that covalent aryl diazonium functionalization suppresses the chemical degradation of exfoliated black phosphorus even following weeks of ambient exposure. This chemical modification scheme spontaneously forms phosphorus-carbon bonds, has a reaction rate sensitive to the aryl diazonium substituent, and alters the electronic properties of exfoliated black phosphorus, ultimately yielding a strong, tunable p-type doping that simultaneously improves field-effect transistor mobility and on/off current ratio. This chemical functionalization pathway controllably modifies the properties of exfoliated black phosphorus, thus improving its prospects for nanoelectronic applications.

---
[1]*Dept. of Materials Science and Engineering, Northwestern University, Evanston, IL 60208*
[2]*Dept. of Chemistry, Northwestern University, Evanston, IL 60208*
[*]e-mail: m-hersam@northwestern.edu.

Black phosphorus (BP) is a layered semiconductor consisting of corrugated planes of phosphorus atoms with strong intra-layer bonding and weak inter-layer interactions. Few-layer BP has a thickness tunable bandgap spanning from a bulk value of 0.3 eV to a monolayer value of ~2 eV in addition to highly anisotropic charge transport and optical response properties,[1-5] which differentiate it from previously studied two-dimensional nanomaterials. Its favorable field-effect hole mobility of ~1,000 $cm^2$ $V^{-1}$ $s^{-1}$ [6-9] exceeds transition metal dichalcogenides[10,11] and makes it attractive for high-performance electronic devices as well as optoelectronic,[12-14] energy storage,[15] and molecular sensing applications.[16] Although black phosphorus is the most thermodynamically stable allotrope of phosphorus, it chemically degrades in the presence of ambient oxygen and water when exfoliated to nanoscale dimensions,[17,18] resulting in the rapid loss of semiconducting properties.[17] Although this high chemical reactivity impedes the use of exfoliated BP in most applications, it also presents an opportunity to investigate the diverse phase space of phosphorus chemistry[19] in a 2D system.

Chemical modification schemes are of broad utility in manipulating the chemical,[20,21] optical,[22] and electronic[23,24] properties of nanomaterials. Aryl diazonium chemistry, which yields covalent carbon bonds at surfaces, has been extensively investigated to this end.[25,26] This chemistry has been found to greatly enhance the photoluminescence emission of carbon nanotubes,[27] produce a bandgap in graphene,[28] enhance solubility,[29,30] and enable the interfacing of dissimilar molecules.[31] Furthermore, the recent report of forming stable P-C bonds between BP and graphite by high energy mechanical milling[15] suggests the possibility of developing carbon-based chemical passivation schemes for BP.

Here, aryl diazonium chemistry is explored as a route to the covalent functionalization of BP. In particular, few-layer BP (~10 nm in thickness) obtained from mechanical exfoliation, which produces the highest performing transistors,[7,12] is chemically modified with 4-nitrobenzene-diazonium (4-NBD) and 4-methoxybenzene-diazonium (4-MBD) tetrafluoroborate salts. Covalent modification is determined to be thermodynamically favorable from density functional theory (DFT) calculations and is verified experimentally with X-ray photoelectron spectroscopy (XPS) and Raman spectroscopy. Following aryl diazonium functionalization, BP exhibits invariant morphology in atomic force microscopy (AFM) measurements after weeks of ambient exposure, thus revealing successful passivation. Charge transport measurements of chemically modified few-layer BP field-effect transistors (FETs) show a controllable p-type doping effect, which simultaneously improves device on/off current ratio and hole carrier mobility. Overall, aryl diazonium chemistry is shown to imbue BP with favorable chemical and electronic properties, thereby bolstering its prospects as a semiconducting 2D nanomaterial.

**Results and Discussion**

*Aryl Diazonium Modification*

Few-layer BP was prepared by mechanically exfoliating bulk BP on Si/SiO$_2$ substrates. Chemical functionalization was performed according to Fig. 1a by immersing the samples in aryl diazonium salt solutions[32] and allowing modification to proceed spontaneously. Following modification, the samples were vigorously agitated in and rinsed with clean acetonitrile to limit the effects of physisorbed molecules.

Plausible product structures obtained from converged DFT calculations are shown in the right inset of Fig. 1a, indicating that chemisorption of aryl groups through the formation of P-C bonds onto a sixteen phosphorus atom supercell is thermodynamically favorable, with two aryl moieties per supercell more favorable than one aryl moiety. The introduction of new covalent bonds, however, appears to cause significant lattice distortion as phosphorus atoms adopt four-coordinate bonding. Previous DFT calculations of the adsorption of nitric oxide on monolayer BP predicted the formation of covalent bonds and similar phosphorus bonding geometry.[16] These results suggest that molecules with dangling bonds (i.e., radical species) can directly bond to BP.

Fig. 1b and Fig. 1c show AFM images of BP modified for 30 min in 10 mM 4-NBD solution. The measured height of the BP flake increases from 7.0 ± 0.3 nm to 8.3 ± 0.4 nm, which is consistent with surface chemical modification. In general, for the aforementioned level of aryl diazonium exposure, an increase of approximately 1.5 nm in BP flake height is observed. Moreover, the surface roughness of the BP flakes increases substantially at higher functionalization levels (see Supplementary Fig. S1), ultimately leading to large edge features relative to the surface of the flake. These increases in flake height and surface roughness agree well with previous reports of aryl diazonium covalent modification of graphene.[28,33]

*Spectroscopic Characterization*

XPS and Raman spectroscopy of BP following a series of reaction times at 10 mM benzene-diazonium concentration were used to determine covalent functionalization. Specifically, the chemical state of modified BP was probed with core-level C 1s and P 2p XPS measurements on separate samples functionalized for increasing reaction times with 4-NBD (Fig. 2a) and 4-MBD (Fig. 2b). For reaction times of 30 min, both 4-NBD and 4-MBD modifications show P-C formation at ~284 eV binding energy in the C 1s spectra.[15] The area of this spectral feature increased for both molecules after 180 min of functionalization. In the P 2p measurements, differences between 4-NBD and 4-MBD emerge. After 30 min, a broad peak at ~133 eV, which agrees well with phosphorus-aryl compounds,[34,35] is evident for 4-NBD, whereas it is only evident after 180 min for 4-MBD. Moreover, the BP doublet at ~130 eV shows a marked decrease in intensity for 4-NBD after 180 min that is more pronounced than the decrease for 4-MBD. Analysis of the P-C related subbands in the C 1s and P 2p spectra also show correlation in spectral weight. Therefore, these subbands are attributed to the formation of phosphorus-aryl covalent bonds. XPS measurements after incremental thermal annealing of aryl diazonium functionalized BP in $N_2$ are shown in Supplementary Fig. S3. Aryl groups are found to persist on BP in excess of the temperature of physisorbed species produced in the aryl diazonium functionalization of graphene,[29] further suggesting a covalent interaction.

Confocal Raman spectroscopy measurements were taken on individual flakes of 10-15 nm in thickness to further elucidate the extent of covalent functionalization of BP. The BP $A^1_g$ mode at 361 cm$^{-1}$ corresponds to primarily out-of-plane atomic displacements.[36] Unlike the $B_{2g}$ and $A^2_g$ modes, the $A^1_g$ mode is nearly constant in intensity when normalized to the Si TO phonon of the substrate.[37] For both 4-NBD (Fig. 2c) and 4-MBD (Fig. 2d), the normalized peak intensity of the $A^1_g$ mode decreased with increasing functionalization. The integrated intensities of these normalized peaks are plotted in Fig. 2d, where it is clear that the $A^1_g$ mode for BP modified with 4-NBD diminishes more rapidly than BP modified with 4-MBD such that 20 hours of reaction with 4-NBD resulted in a loss of BP Raman modes. Furthermore with constant crystallographic orientation, the normalized peak intensities of the $A^1_g$, $B_{2g}$, and $A^2_g$ Raman modes were all found to decrease following aryl diazonium functionalization, as shown in Supplementary Fig. S4.

The weakening of characteristic BP features in XPS and Raman spectroscopy in functionalized samples indicates that intralayer phosphorus bonding is disrupted, and is consistent with DFT calculations of chemisorption of aryl groups producing BP lattice distortion. Furthermore, the observed differing rates of conversion for 4-NBD and 4-MBD functionalization can be attributed to the electronic properties of these molecules. The covalent aryl diazonium reaction mechanism depends on electron transfer from surfaces to the aryl diazonium ions. This electron transfer liberates $N_2$, creating a highly reactive aryl radical that can form a covalent carbon bond to the substrate.[38] Consequently, the electron transfer is the rate limiting step for aryl diazonium reactions, and depends on the overlap of electron states of the targeted substrate and the aryl diazonium ion.[39] Since 4-NBD and 4-MBD have disparate Hammett constants[40] and electrochemical reduction potentials,[41] the lower reduction potential (greater electron richness) of 4-MBD relative to 4-NBD results in less favorable electron transfer, which is manifested in slower P-C bond formation in the XPS and slower reduction of the Raman $A^1_g$ peak intensity.

*Chemical Passivation*

With spectroscopic evidence of covalent BP modification via the formation of covalent P-C bonds, the ambient stability of chemically modified BP was evaluated. Unencapsulated few-layer BP chemically decomposes into phosphorus oxide species when exposed to ambient water and oxygen,[17,18] leading to loss of conduction[17] and the appearance of large morphological protrusions.[9,17] In contrast, exfoliated BP functionalized with aryl diazonium salts exhibit markedly improved morphological stability. The morphology of a thin flake measured by AFM immediately following 30 min functionalization in 10 mM 4-NBD solution is shown in Fig. 3a. After 10 days of exposure to ambient conditions, the flake morphology remains constant with no discernible evidence of oxidation as shown in Fig. 3b. A histogram of the surface roughness of the functionalized flake is presented in Fig. 3c, where there is overlap in the roughness distribution, which is in stark contrast to the degradation of pristine BP flakes. For example, Fig. 3d shows a BP flake with smooth morphology immediately after exfoliation, which evolves into a surface riddled with protrusions after equivalent ambient exposure (Fig. 3e). Consequently, the roughness distributions for the unmodified BP flake shown in Fig. 3f have minimal overlap.

This morphological invariance of 4-NBD modified BP was found to persist for as long as 25 days (Supplementary Fig. S6). After this time, the difference in morphology for functionalized and unmodified flakes becomes accentuated as surface protrusions in degraded BP grow larger and coalesce. Angle-resolved XPS measurements, shown in Supplementary Fig. S9, of functionalized BP after 15 days of ambient exposure have a P-C feature in the C 1s spectra that remains and is localized to the BP surface. Moreover, the BP doublet in the P 2p spectrum remains defined near the sample surface, whereas the unfunctionalized P 2p spectrum has no such evidence of unoxidized BP. The robust morphology and chemical composition of aryl diazonium modified BP demonstrate that covalent P-C bonds can passivate the otherwise highly reactive few-layer BP.

*Controllable p-type Doping of BP FETs*

The effect of aryl diazonium chemistry on the electronic properties of exfoliated BP was investigated with FET device measurements. The transfer characteristics of individual devices were measured with increasing functionalization. Fig. 4a shows transfer curves ($I_{Drain}$ vs. $V_{Gate}$) for BP functionalized with 10 mM 4-NBD. Chemical modification at this concentration produces a strong p-type doping that enhances FET on and off currents, and reduces current modulation as BP becomes degenerately doped. Three hours of aryl diazonium functionalization results in a total loss of conduction that is consistent with the disruption of the BP lattice indicated by XPS and Raman measurements. Analogous observations have been reported for aryl diazonium functionalization of graphene FETs, where doping[33,42] is attributed to noncovalent charge transfer complexes[43] and decreased device current is explained by covalent interactions that disrupt the π-atomic orbitals contributing to conduction.[44] Although the role of noncovalent aryl diazonium interactions cannot be fully discounted, functionalization for 30 min in 10 mM nitrobenzene (Supplementary Fig. S11) resulted in only weak p-type doping. Moreover, the evolution of BP electronic properties correlates with the aforementioned spectroscopic evidence of covalent modification, particularly the disruption of BP bonding at high functionalization. Specifically, the observed electronic property changes are consistent with electrons being transferred from BP during aryl diazonium functionalization, resulting in p-type doping.

The p-type doping effect of aryl diazonium functionalization can be better controlled at low concentrations. In this manner, detrimental increases in the off current can be minimized, in contrast to the charge-transfer p-type doping of BP FETs via $MoO_3$ films.[45] For example, BP functionalized with 1 μM 4-NBD solutions retains high current modulation as shown in Fig. 4b. The associated device metrics are calculated in Fig. 4c and show three distinct regions as a function of the level of BP chemical modification. Initial functionalization (region 1) is marked by p-type doping that simultaneously enhances the hole mobility and on current, leading to higher on/off ratios, which peak and stabilize for intermediate levels of functionalization (region 2). At higher levels of exposure, 1 μM 4-NBD eventually degrades BP FETs (region 3) by first decreasing conduction disproportionately, leading to a high on/off ratio of $10^6$, and culminating in a loss of electrical conduction. Further control of the doping level can be achieved by utilizing the slower reaction kinetics of 4-MBD. The behavior of BP devices modified with 1 μM 4-MBD is similar to 4-NBD (Fig. 4d), with the distinction that devices continue to exhibit enhancement of current after one minute of exposure with the onset of degradation occurring later.

**Conclusion**

Exfoliated BP is spontaneously modified by aryl diazonium molecules forming covalent phosphorus-carbon bonds. The conversion rate is sensitive to the reduction potential of the aryl diazonium molecule, thus allowing control of the level of chemical modification through molecular choice. Low levels of functionalization result in enhanced semiconductor performance, with FETs having improved hole mobility and on/off current ratio. Further functionalization produces passivated BP flakes with stable morphology in AFM following extended exposure to ambient conditions. High levels of functionalization ultimately lead to the loss of BP XPS peaks, Raman modes, and conduction. This aryl diazonium modification strategy allows the properties of BP to be tailored and represents the first investigation of the diverse chemistry of phosphorus applied to a 2D nanomaterial system. Further investigation of the chemistry of BP is likely to yield additional desirable outcomes including the creation of stoichiometric derivatives that exhibit widely tunable chemical, optical, and electronic properties.

**Methods**

**Sample Functionalization.** Few-layer BP was prepared by mechanical exfoliation of bulk BP (Smart Elements) with Scotch Magic Tape on 300 nm thick $SiO_2$/Si substrates (SQI) that were sonicated in isopropanol (ACS reagent grade, BDH) and blow dried with $N_2$. The resulting samples were functionalized in various concentrations of 4-nitrobenzene-diazonium tetrafluoroborate (97%, Aldrich) and 4-methoxybenzene-diazonium tetrafluoroborate (98%, Aldrich) in 100 mM tetrabutylammonium hexafluorophosphate (>99%, Fluka) in acetonitrile (>99.5%, ACS reagent grade) obtained from Sigma-Aldrich for the indicated reaction times. After functionalization, samples were physically agitated in a large volume of neat acetonitrile and blow dried with $N_2$.

**Sample Characterization.** AFM characterization was conducted in noncontact mode on an Asylum Research Cypher atomic force microscope. AFM imaging before and after chemical functionalization was performed in an environmental cell under flow of $N_2$. All other measurements were completed with an ambient scanner. XPS measurements were performed in ultrahigh vacuum in a Thermo Scientific ESCALAB 250 Xi with a nominal spot size of 400 μm. Confocal Raman spectroscopy was measured with a Renishaw InVia Raman Microscope using a 514 nm laser at ~1.38 mW power and a 100x objective with NA of 0.85 on samples capped with a thin layer of $AlO_x$ grown by ALD via the method reported in ref. 17.

**DFT Calculations.** Methods can be found in Supplementary Section S1.

**BP FET Fabrication.** FET devices were fabricated using electron beam lithography on BP samples coated with poly(methyl methacrylate). Electrodes were composed of ~20 nm of Ni and ~40 nm of Au. To minimize BP degradation, lift-off of PMMA occurred in a nitrogen glove box with anhydrous acetone (99.8% Acros Organics). Overall, the exfoliated BP samples spent less than 15 min either unencapsulated by PMMA or in ambient preceding charge transport measurements.

**Acknowledgements**

We acknowledge useful discussions with Joohoon Kang and Xiaolong Liu. This research was supported by the Office of Naval Research (N00014-14-1-0669), the Materials Research Science and Engineering Center (MRSEC) of Northwestern University (NSF DMR-1121262), the National Institute of Standards and Technology (NIST CHiMaD 70NANB14H012), and the Department of Energy (DE-FG02-09ER16109). The work made use of the NUANCE Center, which has received support from the MRSEC (NSF DMR-1121262), State of Illinois, and Northwestern University. Raman spectroscopy measurements were conducted at Argonne National Laboratory's Center for

**Author Contributions**

C.R.R. conceived the experiments with the assistance of J.D.W. and S.A.W., executed the aryl diazonium chemistry, performed the AFM and Raman spectroscopy measurements and analysis, and prepared the manuscript. J.D.W. performed XPS measurements and analysis and contributed to the analysis of AFM and Raman spectroscopy data. S.A.W. fabricated BP FETs and analyzed the device data. Y.Y. conducted DFT calculations and their analysis with the guidance of G.C.S. D.J. assisted with BP FET measurements and analysis. T.J.M. and M.C.H. oversaw the development and execution of the research. All authors contributed to the revision of the manuscript.

**Additional Information**

The authors declare no competing financial interests.

**Figures**

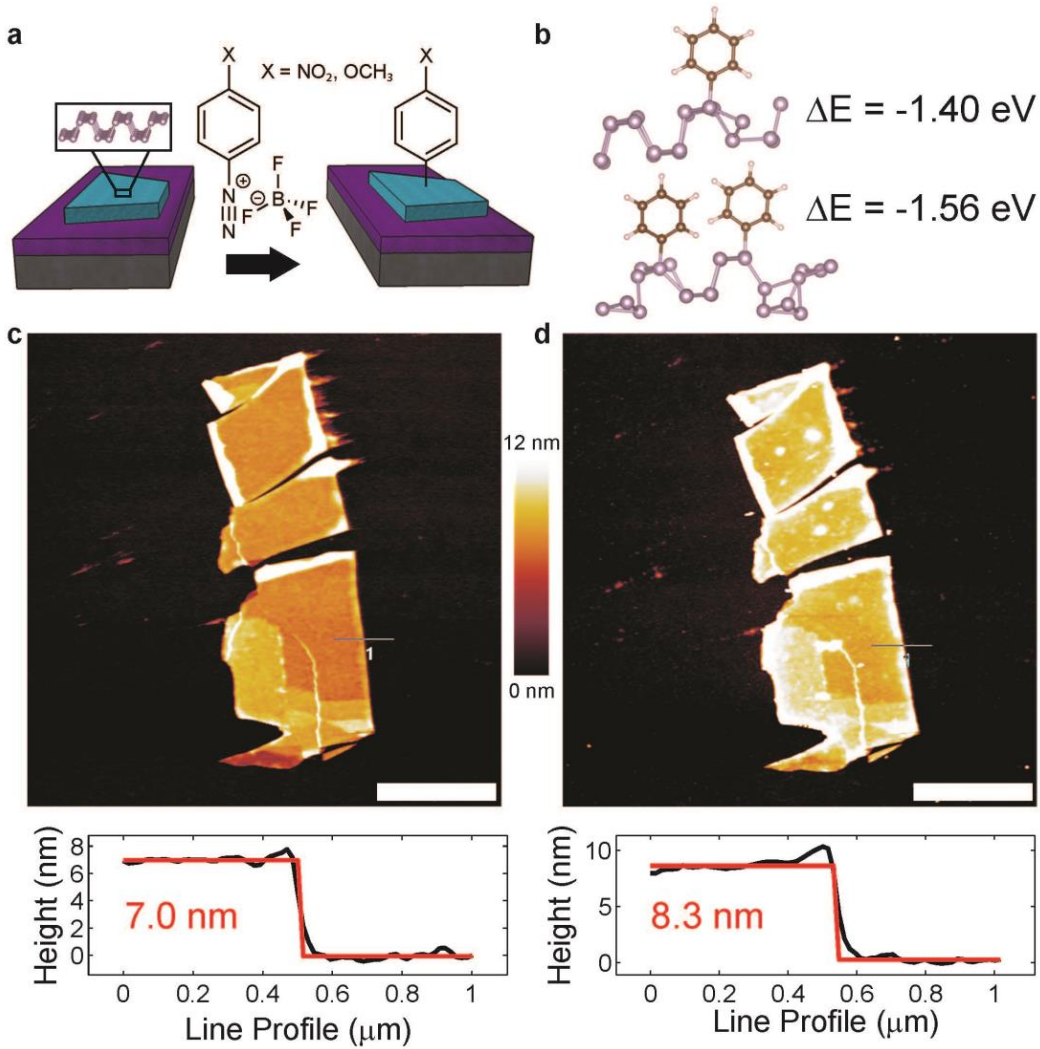

**Figure 1 | Aryl diazonium functionalization of black phosphorus. a,** Reaction scheme of benzene-diazonium tetrafluoroborate derivatives and mechanically exfoliated few-layer black phosphorus (light blue) on a Si (grey)/SiO$_2$ (purple) substrate. The left inset shows the pristine structure of BP while the right inset shows DFT calculated structures of thermodynamically favored covalent bonding of aryl groups to black phosphorus. The adsorption of aryl groups is predicted to produce BP lattice distortion. Carbon atoms – brown, hydrogen atoms – white, phosphorus atoms – purple. The adsorption energy per aryl group is defined as $\Delta E = [E(nC_6H_5/BP) - E(BP) - E(nC_6H_5)]/n$, where $n = 1$ or 2 depending on the number of aryl groups. E($nC_6H_5$/BP) is the energy of covalently bound aryl group(s) to BP, and $E$(BP) and $E(nC_6H_5)$ are the initial energies of unassociated BP and aryl molecule(s), respectively. **b,** AFM micrograph of a black phosphorus flake prior to functionalization with flake height profile. **c,** AFM micrograph (top) of the same flake following 30 min exposure to 10 mM 4-NBD with height profile (bottom) showing an increase in flake height that is attributed to the attachment of aryl groups. AFM scale bars are 2 µm.

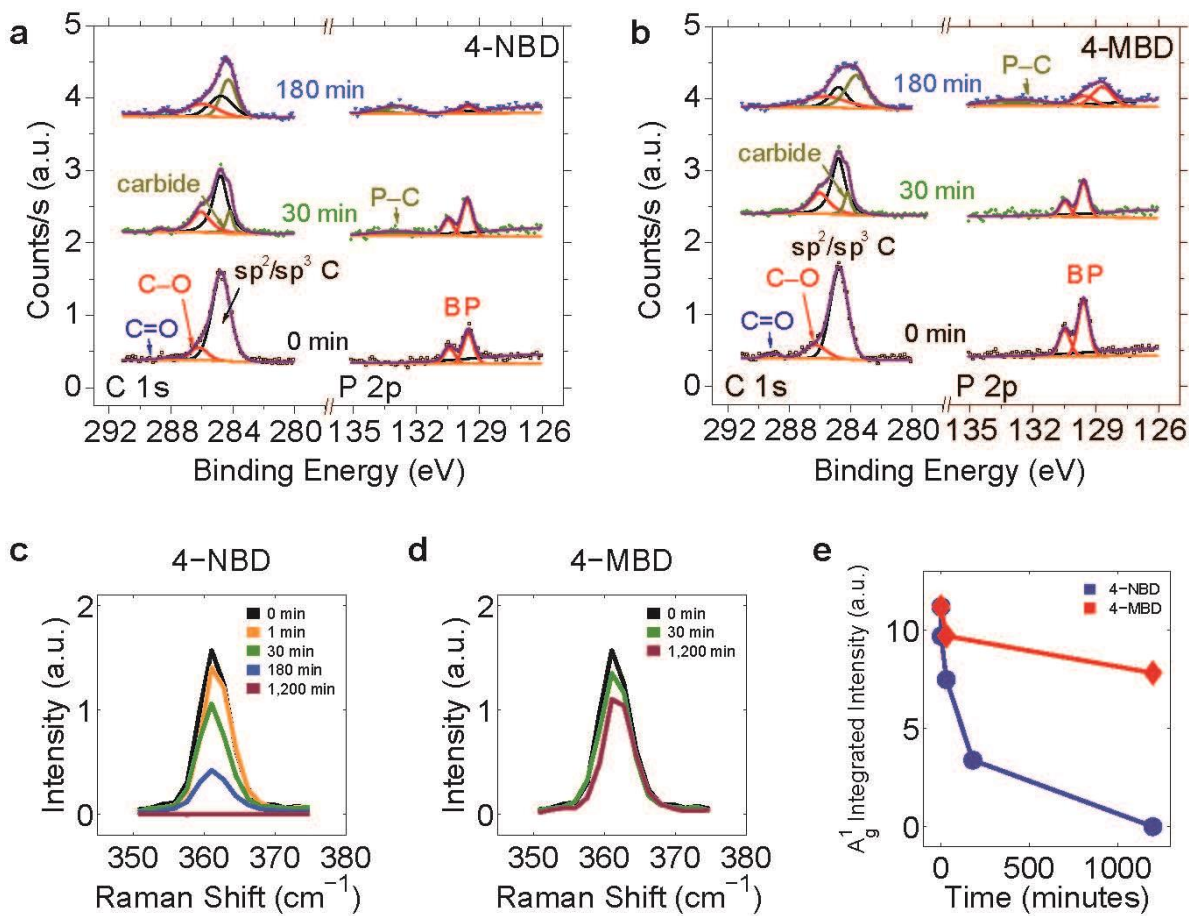

**Figure 2 | Spectroscopic characterization of BP with increasing 10 mM aryl diazonium functionalization. a,** Deconvoluted C 1s and P 2p core level XPS spectra of 4-NBD functionalized BP. After 30 min, a P-C bonding feature (distinct from adventitious carbon) is apparent at 284 eV in the C1s spectrum, and increases in spectral weight after 3 hr. In the P 2p spectrum, a broad P–C feature at 133 eV is evident after 30 min, and grows in intensity after 3 hr as the BP doublet at 130 eV diminishes. **b,** The corresponding XPS spectra for 4-MBD. The C1s spectrum shows similar P-C formation after 30 min and 3 hr. In the P 2p spectrum, P–C bonding is apparent after 3 hr while the BP doublet remains prominent, although broadened, indicating slower reaction kinetics. P 2p spectra are multiplied by 10 for visual clarity. **c,** The BP $A^1_g$ Raman mode diminishes with increasing exposure to 4-NBD. **d,** The $A^1_g$ mode shows a similar trend for 4-MBD, albeit at a significantly reduced rate. **e,** Integrated $A^1_g$ intensities obtained from (**c**) and (**d**) illustrating the different rate of electron transfer from BP for 4-NBD and 4-MBD. Raman intensities are normalized to the Si substrate TO phonon peak intensity.

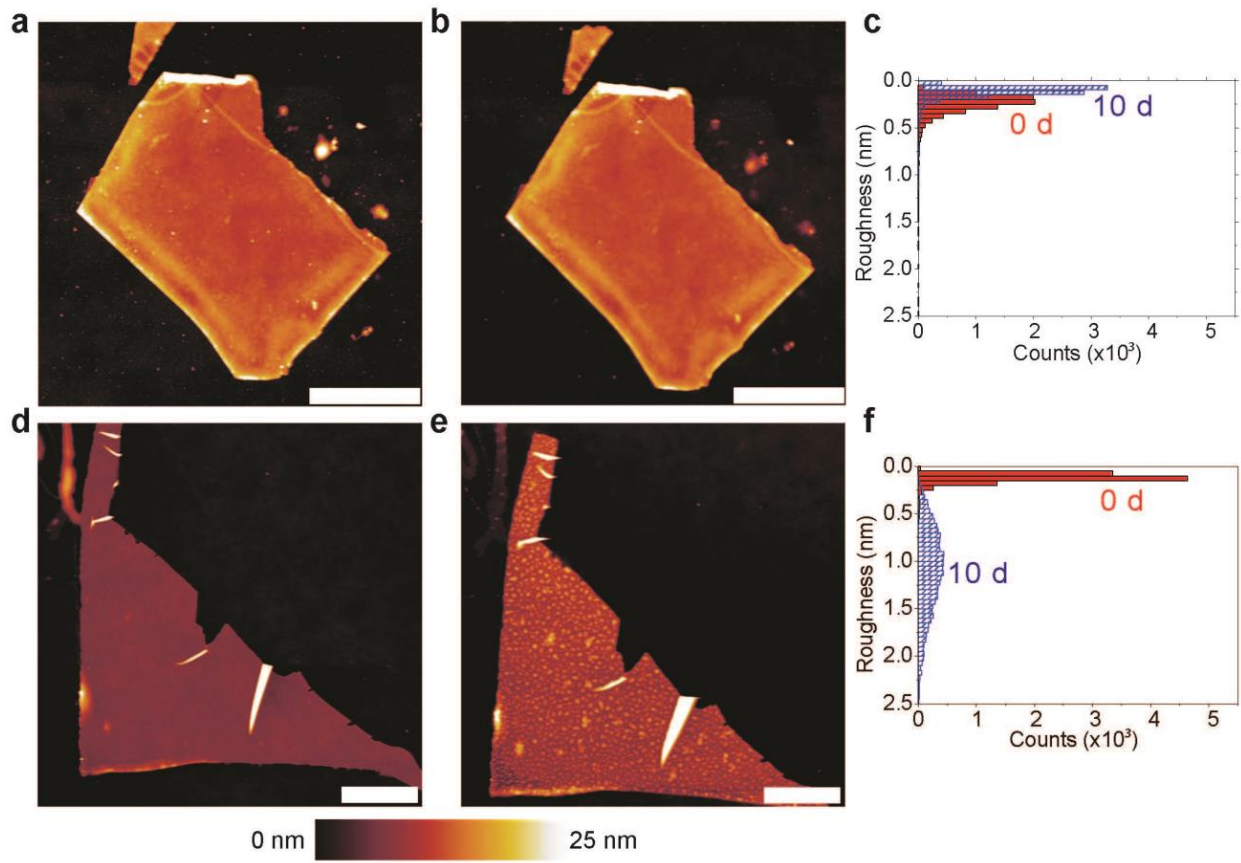

**Figure 3 | Chemical passivation of black phosphorus.** AFM characterization of BP morphology before and after ambient exposure. **a,** BP flake immediately after functionalization with 10 mM 4-NBD for 30 min. **b,** The same flake as in (**a**) after 10 days of ambient exposure. **c,** Histograms of the surface roughness of the flakes in (**a**) and (**b**) with overlap in the distributions denoted with purple color. **d,** Pristine BP flake immediately after exfoliation. **e,** The same flake in (**d**) after 10 days of ambient exposure with distinct morphological protrusions indicative of chemical degradation. **f**, Histograms of the surface roughness of the flakes in (**d**) and (**e**), demonstrating the large change in surface roughness after ambient exposure. AFM scale bars are 2 μm.

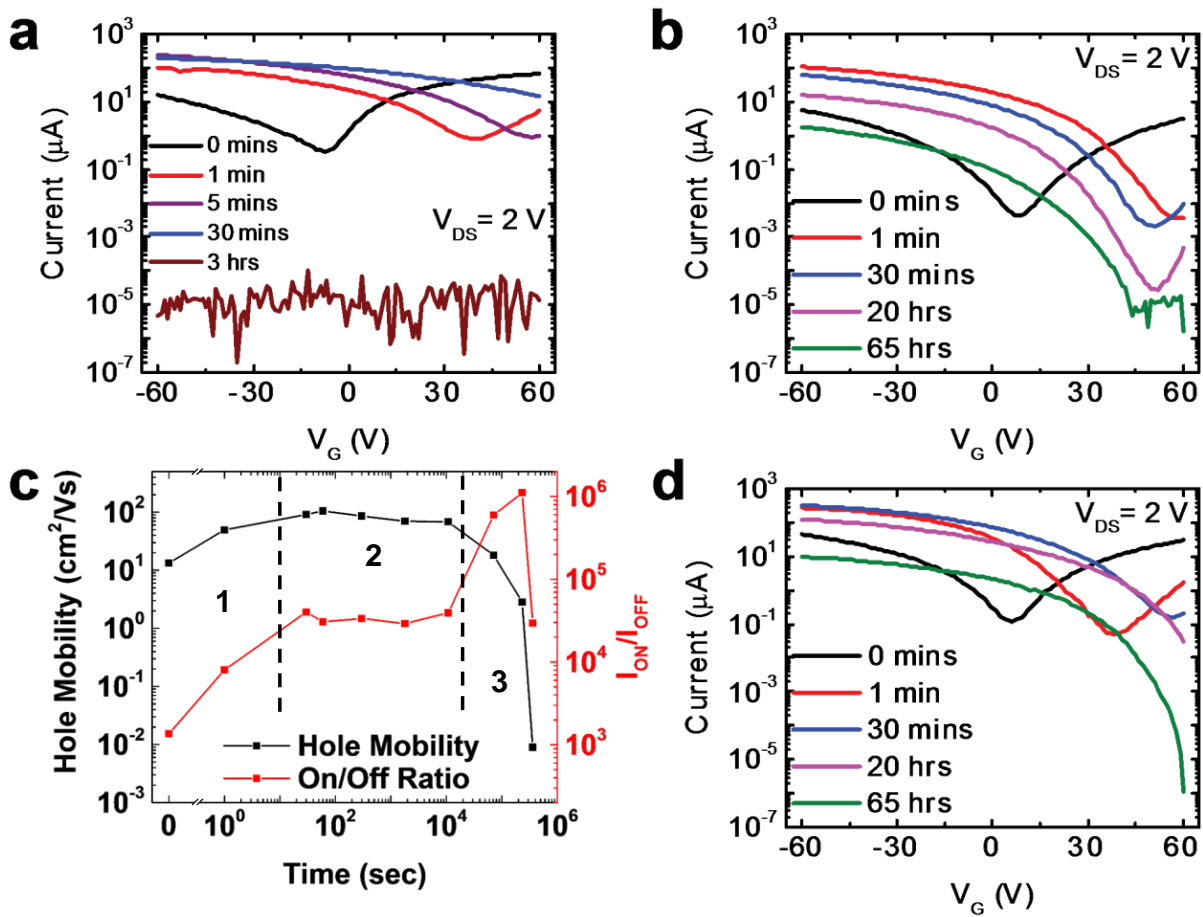

**Figure 4 | Effect of covalent functionalization on charge transport in exfoliated black phosphorus field-effect transistors. a,** Transfer curves for an exfoliated BP FET as a function of exposure time to 10 mM 4-NBD, demonstrating strong p-type doping and loss of conduction. **b,** Transfer curves for an exfoliated BP FET as function of exposure time to 1 μM 4-NBD. **c,** Hole mobility and on/off current ratio for the exfoliated BP FET from (**b**) as a function of exposure time to 1 μM 4-NBD, showing three regions corresponding to: (1) increased p-type doping, (2) performance that is relatively independent of exposure time, and (3) degradation. **d,** Transfer curves for an exfoliated BP FET as a function of exposure time to 1 μM 4-MBD, showing slower reaction kinetics.

# Supplementary Information

# Covalent Functionalization and Passivation of Exfoliated Black Phosphorus via Aryl Diazonium Chemistry

Christopher R. Ryder[1], Joshua D. Wood[1], Spencer A. Wells[1], Yang Yang[2], Deep Jariwala[1], Tobin J. Marks[1,2], George C. Schatz[2] and Mark C. Hersam[1,2*]

**Table of Contents**

**Density functional theory calculation methods**

Calculations were performed using the VASP package.[1,2] The exchange-correlation potential was described by the generalized gradient approximation (GGA) using the PBE functional.[3,4] An energy cutoff of 400 eV was used for the plane-wave basis set. The conjugate gradient method was used to optimize the ion positions.[5] Valence electrons included in the calculations were P $3s^23p^3$, C $2s^22p^2$ and H $1s^1$, and the interaction between ions and electrons was described by the projector augmented wave (PAW) method.[6,7]

**Atomic force microscopy (AFM) characterization for increasing aryl diazonium functionalization**

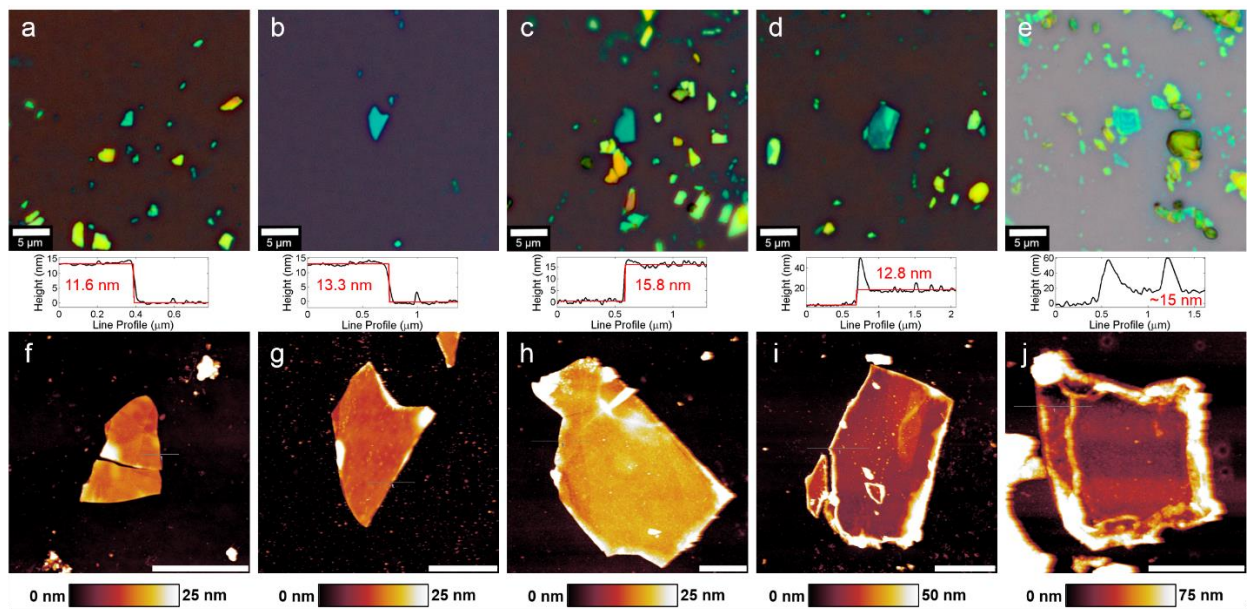

**Figure S1 | Evolution of flake morphology with increasing 10 mM 4-NBD functionalization. a-e,** Optical micrographs at 100× magnification. **a,** 0 min (pristine). **b,** 1 min. **c,** 30 min. **d,** 3 hr. **e,** 20 hr. Tall edges are apparent after 20 hr of reaction. **f-k,** AFM measurements of flake morphology corresponding to the same flakes in (**a-e**) with the flake height profile. The edge features show increased height at 3 hr and 20 hr. These flakes were measured with confocal Raman spectroscopy in Fig. 2c and used to quantify the amount of functionalization in Fig. 2e. Samples have been capped with a ~2 nm thick AlO$_x$ layer grown by atomic layer deposition to allow for laser irradiation.

**AFM of flake before and after being submerged in electrolyte and solvent for 30 min**

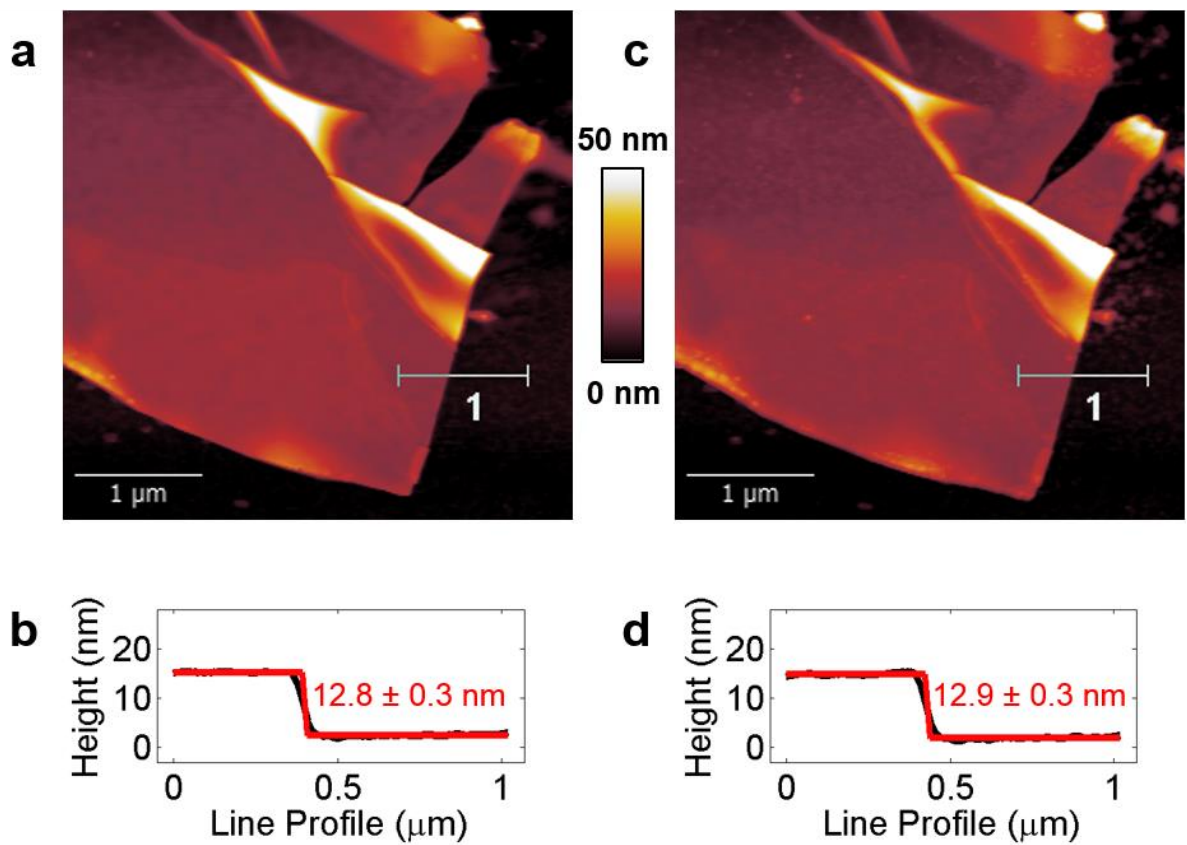

**Figure S2 | Invariant flake height before and after submersion in 100 mM NBu$_4$PF$_6$ in acetonitrile for 30 min. a,** AFM micrograph of a BP flake taken immediately following exfoliation in an environmental cell under laminar flow of nitrogen. **b,** Height profile of the line indicated in **(a)**. The flake height is initially measured to be 12.8 ± 0.3 nm. **c,** AFM micrograph of the same flake immediately following submersion in a 100 mM NBu$_4$PF$_6$ solution in acetonitrile for 30 min and following the same rinsing procedure used for aryl diazonium functionalization. This image was acquired in the same environmental cell under the same conditions. **d,** Height profile of the line indicated in **(c)**. The flake height measured after submersion in the electrolyte and solvent is 12.9 ± 0.3 nm, showing negligible change in contrast to measurements of BP flakes exposed to aryl diazonium solutions.

# X-ray photoelectron spectroscopy (XPS) study of the thermal stability of aryl diazonium modification

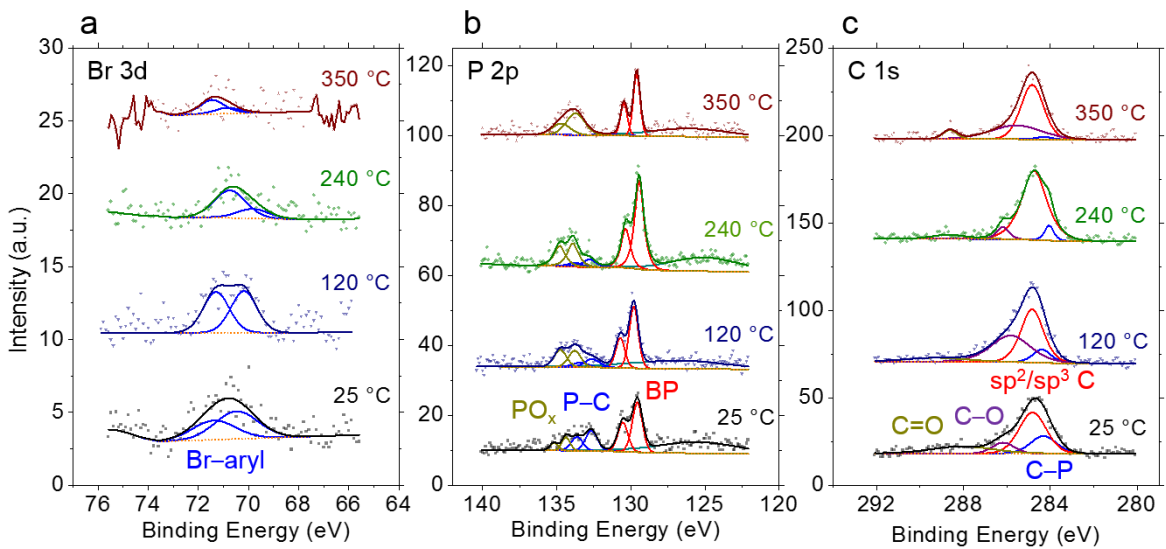

**Figure S3 | Temperature dependent X-ray photoelectron spectra for 4-bromobenzene-diazonium tetrafluoroborate (4-BBD) functionalized BP**. All heating steps were performed for 10 min *ex situ* in a $N_2$ environment (temperature setpoint ±1°). BP was functionalized with 10 mM 4-BBD with 100 mM $NBu_4PF_6$ in acetonitrile. **a,** Br 3d core level spectra for functionalized BP at room temperature (~25 °C, black), after 120 °C heating (navy), after 240 °C heating (green), and after 350 °C heating (maroon). A doublet corresponding to Br attached to an aryl ring (Br–aryl, ~70.5 eV)[8] is measured at room temperature following functionalization, and persisted after annealing at 120 °C and 240 °C. After annealing at 350 °C, residual C–Br traces (~70.9 eV) remained. **b,** P 2p core level spectra for the same functionalization and heating conditions as **(a)**. At room temperature, both carbide (P–C, blue, ~132.7 eV) and oxide ($PO_x$, gold, ~134.4 eV) moieties[9,10] appear with the BP doublet (~129.7 eV).[9] The existence of distinct P–C bonds and Br–aryl groups confirms the covalent grafting of brominated aryl groups to BP. As the sample is heated at 120 °C, 240 °C, and 350 °C, the P–C subbands decrease in quantity and eventually are lost, leading to increased BP oxidation. **c,** C 1s core level spectra for the same functionalization and annealing conditions as **(a)** The emergence and loss of the carbide (C–P) band (blue, ~284.1 eV) with respect to increasing temperature tracks the P–C doublets of **(b)** and the Br–aryl groups of **(a)**. Functionalized BP regions are removed after ≥240 °C treatment, with some trace bromine species remaining after annealing at 350 °C, which is above the evaporation temperature of physisorbed species produced from the aryl diazonium functionalization of graphene.[11]

**Raman spectra of aryl diazonium functionalized BP**

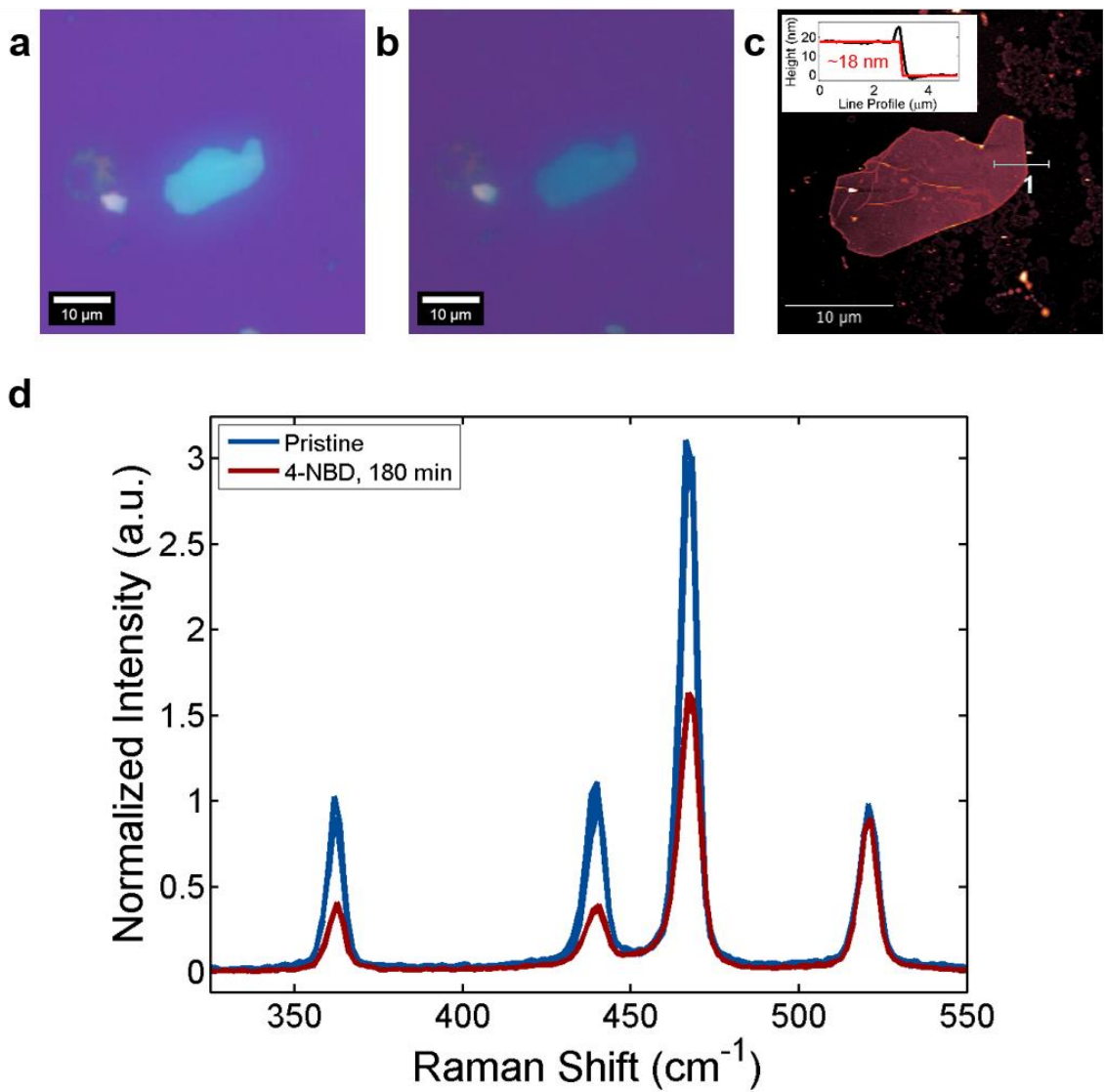

**Figure S4 | Raman spectra of large lateral size 18 nm flake before and after 180 min functionalization in 10 mM 4-NBD. a,** Optical image of a flake immediately following exfoliation taken in a nitrogen environment. **b,** Optical image of the same flake following 180 min functionalization in 10 mM 4-NBD showing the same flake in nearly identical orientation. **c,** AFM of the flake used to determine its thickness, taken after functionalization. **d,** Normalized Raman spectra immediately following exfoliation (blue) and immediately following aryl diazonium functionalization (red). The spectra were acquired in a nitrogen environment in a Linkam $N_2$ environmental cell to prevent ambient degradation and photooxidation during measurements using a 50× long-working distance objective (NA = 0.5). Three spectra are depicted for each case to show the reproducibility of the measurements. Spectra are normalized to the Si TO phonon mode.

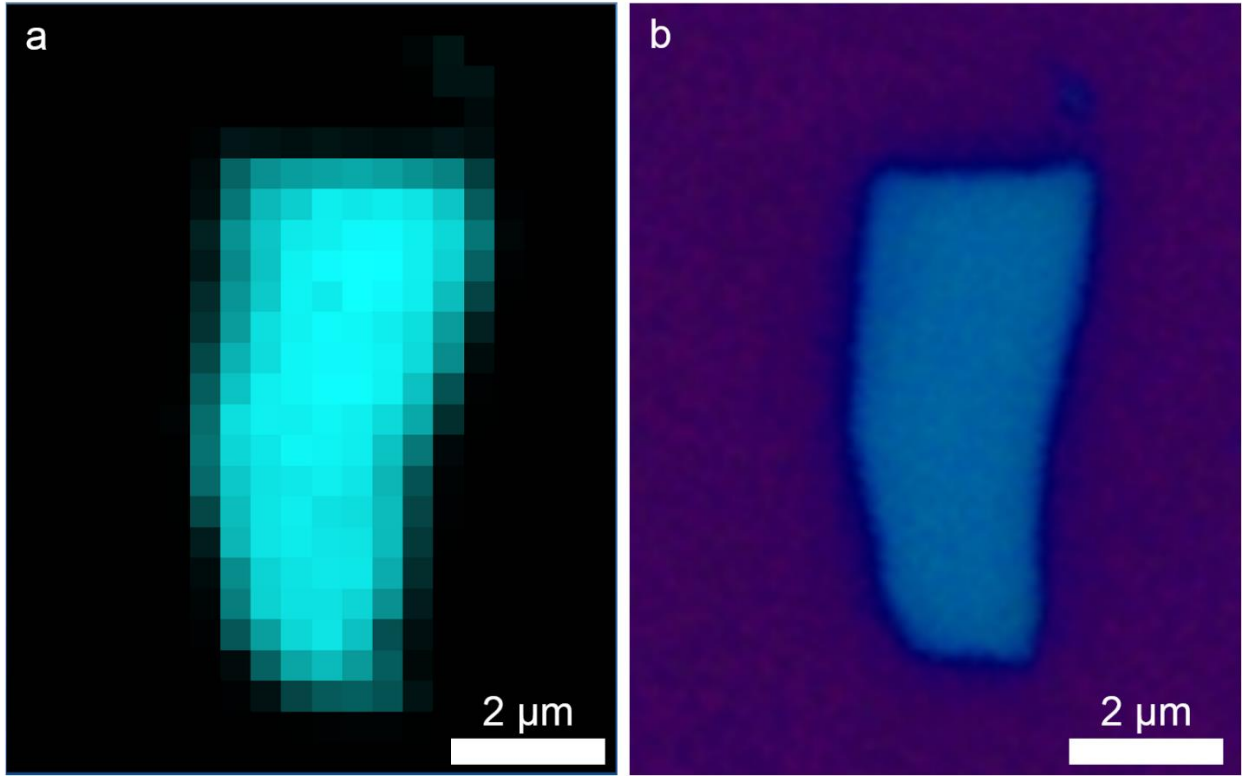

**Figure S5 | Raman mapping of the normalized black phosphorus $A^1_g$ mode following 30 min of 10 mM 4-NBD functionalization. a,** Raman map of the $A^1_g$ mode intensity located at 361 cm$^{-1}$ normalized to the intensity of the Si TO phonon mode of the substrate at 519 cm$^{-1}$. The BP flake is measured in 400 nm increments, and shows uniform normalized Raman intensity. **b,** The corresponding optical image of the measured BP flake.

**AFM characterization of BP after ambient exposure**

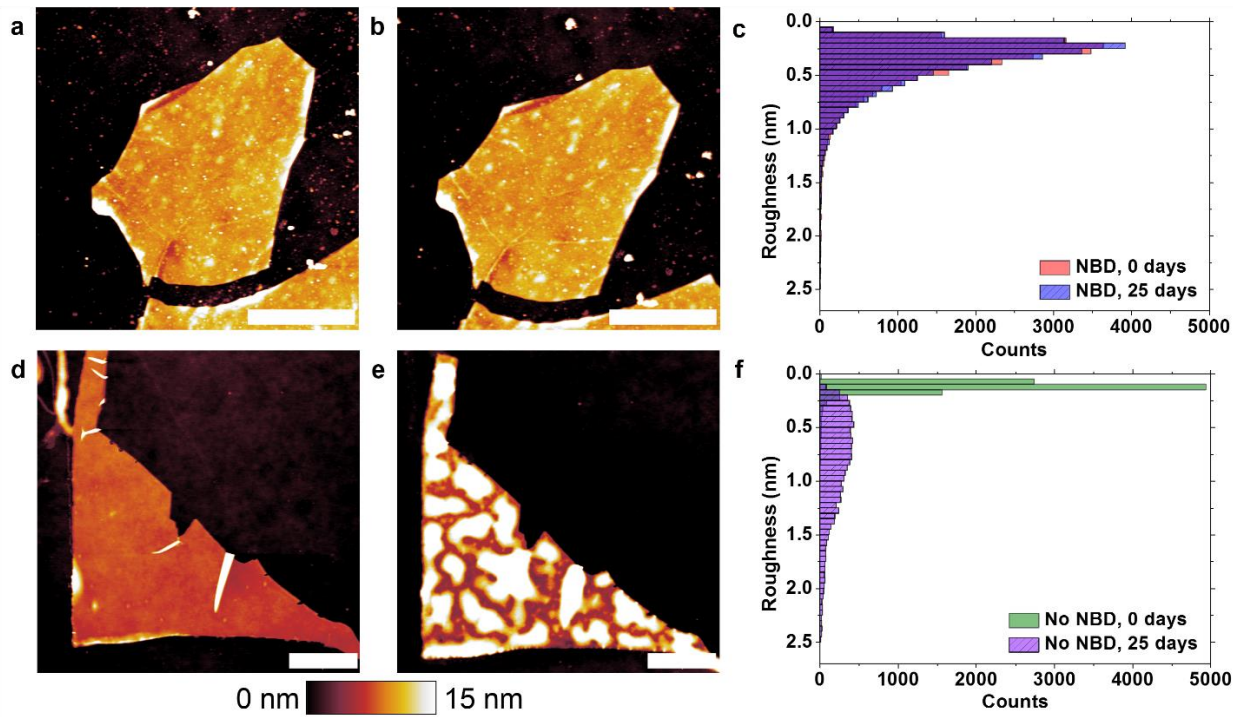

**Figure S6 | AFM of black phosphorus flakes following 25 days of ambient exposure.** AFM characterization of BP morphology before and after ambient exposure. **a,** BP flake immediately after 30 min of functionalization with 10 mM 4-NBD. **b,** The same flake as in (**a**) after 25 days of ambient exposure. **c,** Histograms of the surface roughness of the flakes in (**a**) and (**b**) with overlap in the distributions denoted with purple color. **d,** Pristine BP flake immediately after exfoliation. **e,** The same flake as in (**d**) after 25 days of ambient exposure with distinct morphological protrusions indicative of chemical degradation. **f**, Histograms of the surface roughness of the flakes in (**d**) and (**e**), demonstrating the large change in surface roughness after ambient exposure. AFM scale bars are 2 μm.

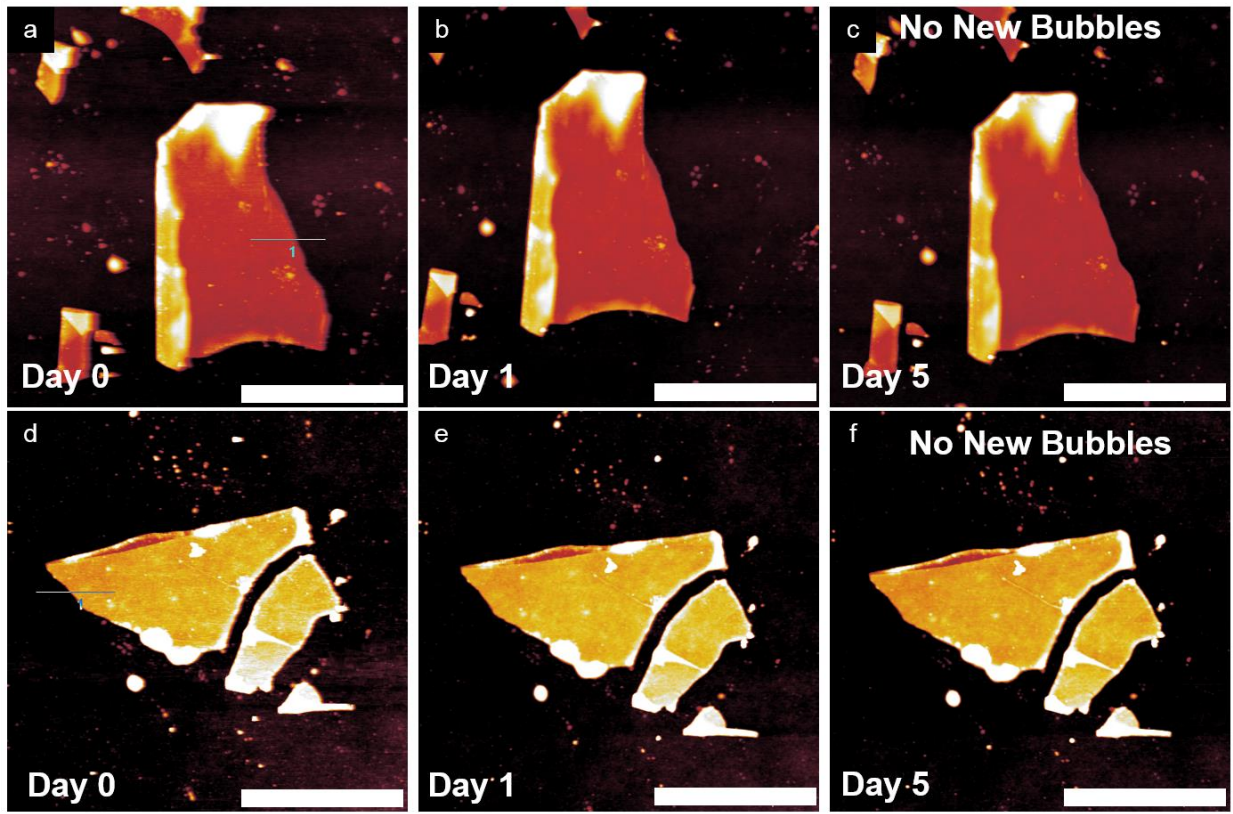

**Figure S7 | Chemical passivation of BP after 1 min of 10 mM 4-NBD functionalization. a-c,** Stable AFM flake morphologies are obtained under these functionalization conditions for a period of five days. **d-f,** The passivation effect is common for flakes that are <20 nm in thickness. All scale bars are 2 μm.

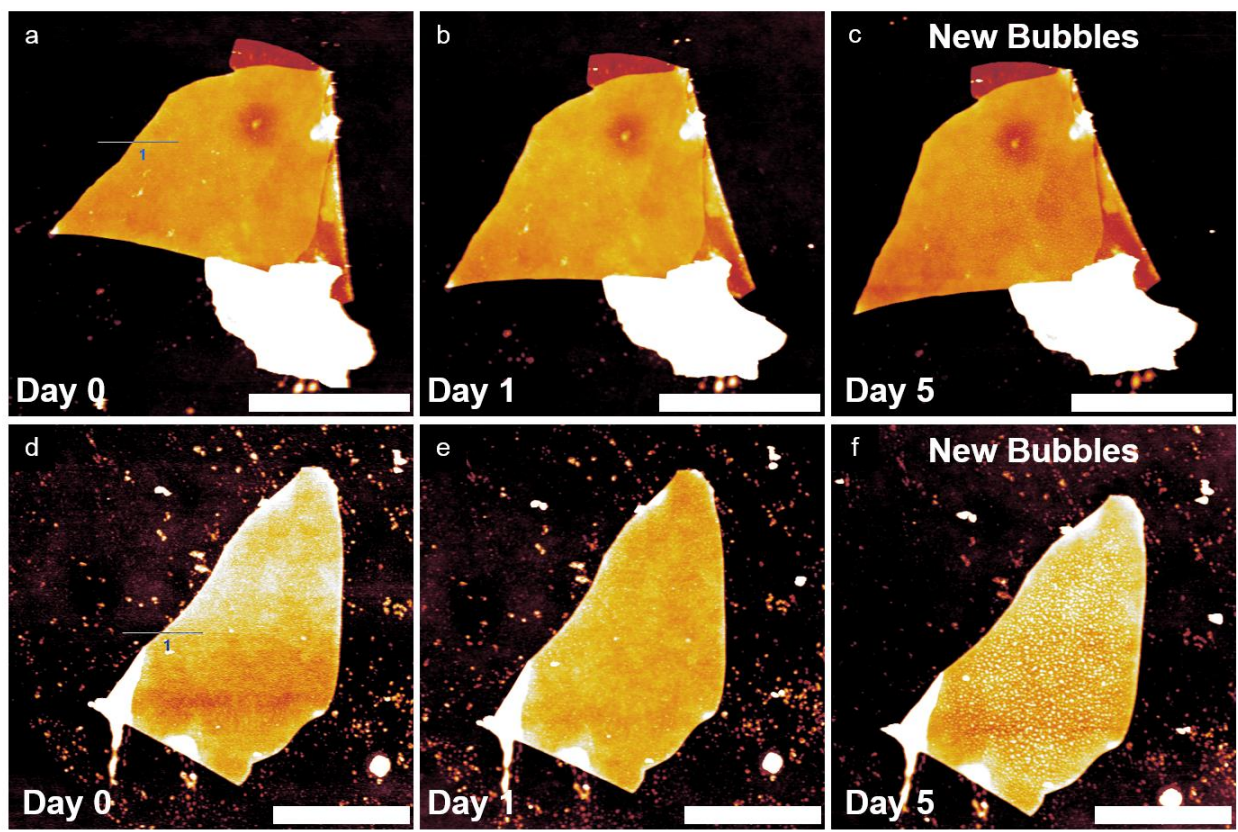

**Figure S8 | Control experiment for probing degradation in BP. a-c,** Exfoliated BP exposed to acetonitrile and tetrabutylammonium hexafluorophosphate exhibits stable AFM morphology for one day in ambient conditions. However, flakes begin to show evidence of morphological protrusions after five days. **d-f,** Another flake showing the same trend over five days of exposure to ambient conditions. All scale bars are 2 μm.

**Angle-resolved XPS of functionalized BP flakes after 15 days of ambient exposure**

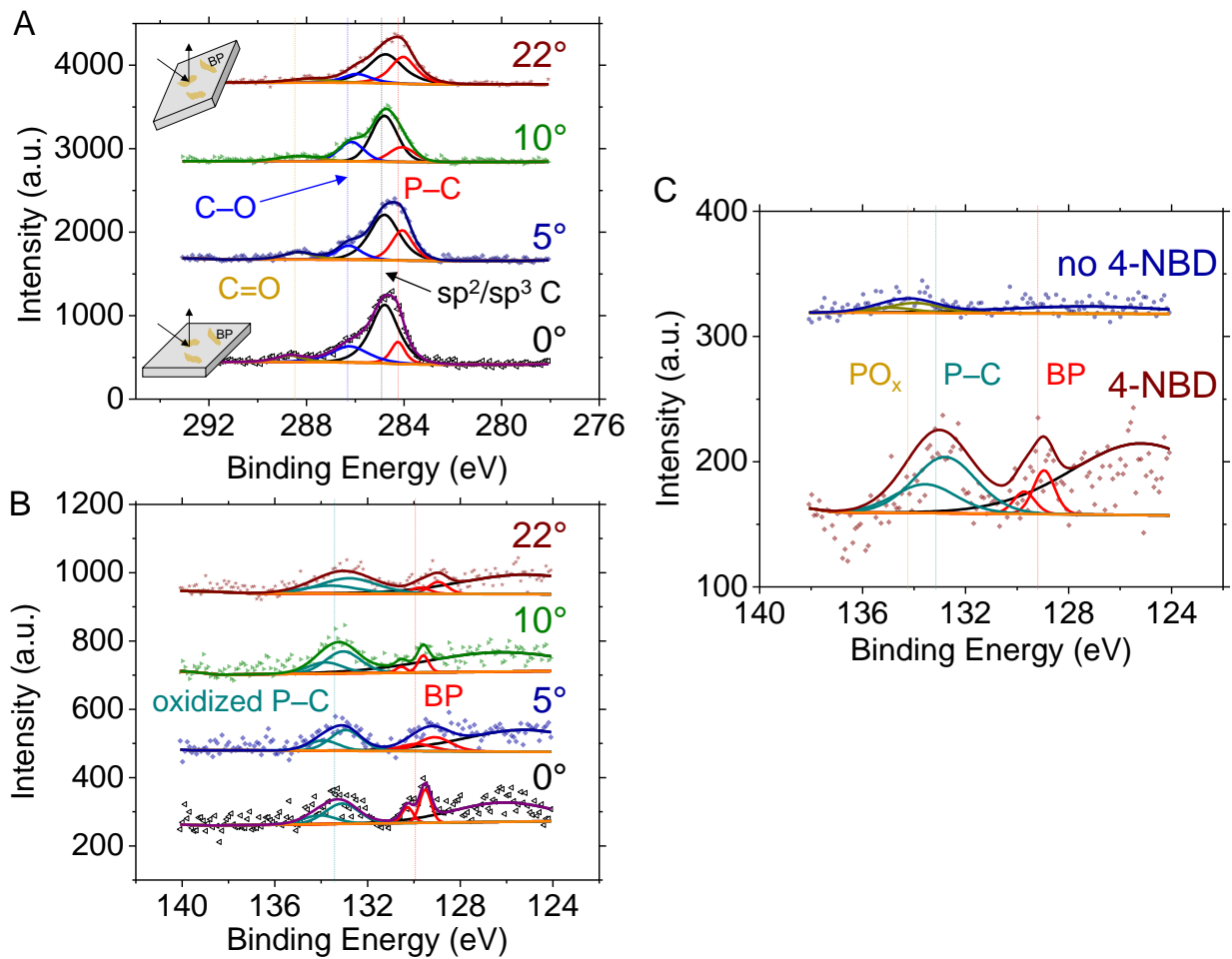

**Figure S9 | Angle-resolved XPS characterization of chemical state after 15 days of ambient exposure.** Samples were functionalized with 10 mM 4-NBD for 30 min and stored in ambient for 15 days. **a,** Angle-resolved C 1s core-level spectra. *Inset*: Schematics showing the sample orientation at 0° and 22° sample angle. With increasing sample angle, the carbide (P–C, red) subband contribution increases relative to the adventitious $sp^2/sp^3$ carbon (black), carboxyl (blue), and carbonyl (gold) subbands. This observation implies that the diazonium functionalization is localized to the sample surface. **b,** Angle-resolved P 2p core-level spectra for the same samples and spots as **(a)**. Oxidized P–C (dark cyan) and BP (red) subbands appear for all angles. The BP subband persists at a high sample incidence angle (22°), suggesting preserved, sub-surface BP below the diazonium functionalized surface. Additionally, the diazonium surface functionalization is confirmed from the increasing oxidized P–C subband area with respect to sample angle. **c,** Surface sensitive, P 2p core-level XPS data for functionalized and unfunctionalized BP. The 4-NBD functionalized sample shows passivation of the sub-surface BP (red subband at ~129.5 eV) and ambient oxidization of the functionalized surface (dark cyan P–C subband at ~133 eV). Conversely, the unfunctionalized BP sample surface and sub-surface are oxidized (gold $PO_x$ subband at ~134 eV).

**Contact angle measurements of aryl diazonium modified BP**

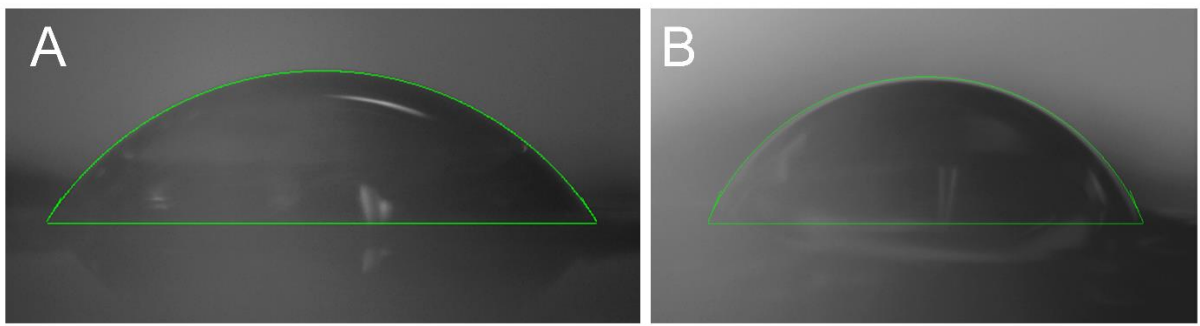

**Figure S10 | Water contact angle measurements of BP aryl diazonium functionalization.** All contact angles fitted by the LBADSA technique and conducted with argon-sparged $H_2O$ droplets. The errors are the standard deviation of multiple measurements. Images of **a,** freshly cleaved BP ($\theta = 56.7 \pm 2.9°$) and **b,** 30 min 10 mM 4-NBD functionalization of BP ($\theta = 63.9 \pm 5.0°$). The production of a more hydrophobic surface on BP may aid in the passivation effect.

**Table S1 | Contact angle measurements of BP aryl diazonium functionalization.**

| Substrate | Contact Angle (deg) |
|---|---|
| Kapton tape | 90.0° |
| Evaporated Au | 70.1° |
| BP, freshly cleaved | 56.7 ± 2.9° |
| BP, 30 min 10 mM 4-NBD | 63.9 ± 5.0° |

## Control studies of BP field-effect transistors

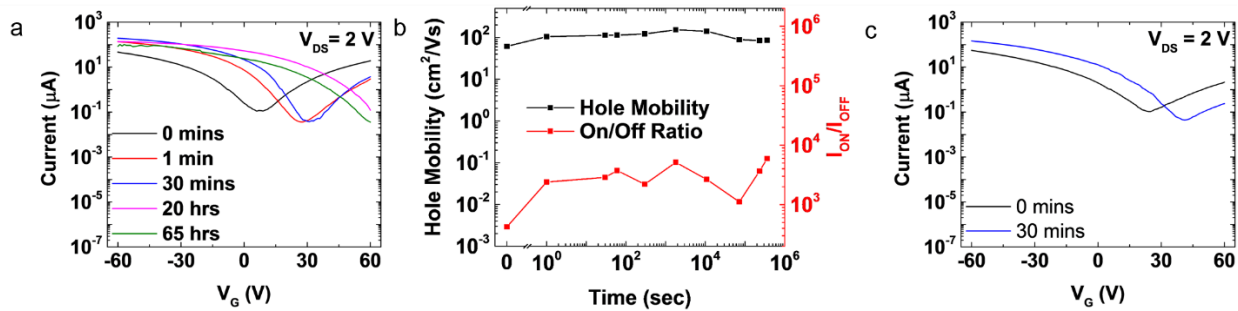

**Figure S11 | Control exfoliated BP field-effect transistors. a,** Transfer curve ($I_D$ versus $V_G$) of an exfoliated BP FET following increasing exposure to acetonitrile and 100 mM tetrabutylammonium hexafluorophosphate showing weak p-type doping. **b,** Evolution of performance metrics for the device in (**a**) showing unaffected hole mobility and relatively constant on/off current ratio. **c,** Functionalization of an exfoliated BP FET with 10 mM nitrobenzene and 100 mM tetrabutylammonium hexafluorophosphate that mimics the noncovalent charge transfer from the aromatic and nitro moiety of 4-NBD. Nitrobenzene modification produces different behavior than that of 4-NBD and resembles that of solvent and electrolyte effects.